\documentclass[12pt, a4paper]{article}
\usepackage[utf8]{inputenc}
\usepackage[T1]{fontenc}
\usepackage{lmodern}  
\usepackage{graphicx}
\usepackage{physics}
\usepackage{braket}
\usepackage{amsmath, amssymb}
\usepackage{geometry}
\usepackage{setspace}
\usepackage{caption}
\usepackage{subcaption}
\usepackage{hyperref}
\usepackage{xcolor}
\usepackage{ulem}
\usepackage{mathtools}
\usepackage{tikz}
\usepackage{bbm}
\usepackage{pdfpages} 

\newsavebox{\mstrut}

\newcommand{\scom}[1]{%
    \sbox{\mstrut}{\(#1\)}%
    \mathinner{\left[\mkern-2mu\left[{#1}\right]\kern-0.1\ht\mstrut\right]}%
}

\title{The Spectrum Lies:\\
Level Repulsion in Open Quantum Systems Lacks Physical Meaning}
\author{
  Caio B. Naves$^{1,*}$, Thomas Klein Kvorning$^{2}$, Jonas Larson$^{1}$ \\
  \\
  {\small $^1$Stockholm University, AlbaNova University Center, 106 91 Stockholm, Sweden} \\
  {\small $^2$Royal Institute of Technology, AlbaNova University Center, 106 91 Stockholm,
Sweden} \\
  {\small $^*$Corresponding author: caio.naves@fysik.su.se}
}
\date{}  

\begin{document}

\maketitle

\begin{abstract}
In Hamiltonian systems, level statistics provide a faithful diagnostic of quantum chaos. 
By analogy, Lindbladian spectral statistics are widely used for open quantum systems, following the Grobe–Haake–Sommers conjecture that systems with chaotic classical counterparts exhibit level repulsion.
Here we identify a fundamental limitation of this analogy. 
Unlike Hamiltonian spectra, the Lindbladian spectral bulk contributes little to long-time dynamics, which are governed by modes with real parts closest to zero. 
Bulk modes typically decay on microscopic time scales unless concentrated near the imaginary axis.
Consequently, perturbations must influence the dynamics before these modes decay to produce observable effects. 
Yet level statistics resolve much finer spectral scales, allowing perturbations to alter spectral correlations without appreciably changing evolution.
Strong Lindbladian non-normality amplifies this separation, enabling drastic changes in level statistics under dynamically negligible perturbations. 
We demonstrate examples exhibiting Ginibre level repulsion without signatures of dynamical chaos at any time.
Moreover, we construct pairs of Lindbladians whose dynamics become indistinguishable for all observables and times in the thermodynamic limit, despite displaying contrasting spectral statistics, with and without level repulsion.
Our results challenge Lindbladian bulk statistics as universal chaos diagnostics, suggesting possible reliability only for approximately normal generators with spectra concentrated near the imaginary axis.
\end{abstract}
\newpage

Already in the early quantum theory of Bohr, Sommerfeld, and others—predating the modern formulations of Schrödinger and Heisenberg—concerns were raised about the compatibility between classical and quantum descriptions of motion. In his 1917 paper~\cite{einstein1917quantensatz}, Einstein questioned whether quantization could be consistently applied to classically non-integrable systems that do not admit global action–angle variables. Decades later, the field of quantum chaos emerged with the goal of understanding how signatures of classical chaotic dynamics—rooted in nonlinear equations of motion—manifest in an underlying quantum theory that is fundamentally linear~\cite{haake1991quantum,gutzwiller2013chaos}. Berry emphasized that, strictly speaking, ``quantum chaos'' does not exist; rather, one can only study how classical chaos imprints itself on quantum systems, a viewpoint he termed quantum chaology~\cite{berry1989quantum}. Despite the linearity of quantum mechanics, numerous remnants of classical chaos persist, including quantum scars~\cite{heller1984bound}, semiclassical Lyapunov-type growth observed in quantities such as the Loschmidt echo and out-of-time-ordered correlators (OTOCs)~\cite{larkin1969quasiclassical,peres1984stability,jalabert2001environment,maldacena2016bound}, and ergodic eigenstates in quantum systems whose classical counterparts are chaotic~\cite{deutsch1991quantum,srednicki1994chaos}.

Inspired by Wigner’s work on random matrices~\cite{wigner1958distribution,wigner1993characteristic}, it was conjectured that the spectra of quantum Hamiltonians exhibit universal statistical features reflecting the nature of the underlying classical dynamics~\cite{berry1977level,bohigas1984characterization,haake1991quantum}. In systems admitting a well-defined classical limit, this connection is encapsulated in the Berry-Tabor and Bohigas--Giannoni--Schmit conjectures: classically regular dynamics leads to uncorrelated (Poissonian) spectra, whereas classically chaotic dynamics gives rise to strongly correlated eigenvalues with level repulsion described by random matrix theory.

However, many quantum systems of current interest—such as strongly interacting lattice models, spin systems, and open quantum systems—either lack a clear semiclassical limit or are intrinsically quantum in nature. In these settings, spectral statistics such as level repulsion, spectral rigidity, and eigenvector delocalization are employed as intrinsic diagnostics of quantum chaos and complex quantum dynamics~\cite{bhattacharyya2022towards}, which is further characterized using dynamical tools such as out-of-time-ordered correlators (OTOCs)~\cite{peres1984stability,jalabert2001environment,maldacena2016bound}, spectral form factors~\cite{bertini2018exact,xu2021thermofield}, quantum complexity~\cite{brown2018second}, operator growth~\cite{hashimoto2023krylov,erdmenger2023universal}, or entanglement scrambling~\cite{swingle2018unscrambling}. This extends Berry’s notion of quantum chaos beyond the semiclassical correspondence and frames it instead as a structural property of quantum many-body dynamics.

Open and dissipative systems introduce an additional layer of complexity. Dissipation contracts the effective phase-space volume, potentially competing with chaotic spreading. Nevertheless, classical dissipative systems can exhibit transient chaos with positive Lyapunov exponents~\cite{bergamasco2023quantum,garcia2024lyapunov}, as well as long-time chaotic dynamics characterized by strange attractors~\cite{brun1996quantum,guckenheimer2013nonlinear,ferrari2025dissipative,stitely2022lasing,rufo2025quantum}. In quantum systems, the dynamics is generated by a superoperator, typically of Lindblad form. In the absence of symmetries or conserved quantities, the Lindbladian superoperator (also referred to as the Liouvillian) generically possesses a unique steady state $\hat{\rho}_\mathrm{ss}$ toward which all initial states relax. Consequently, the asymptotic dynamics is trivial in the sense that the system ultimately loses memory of its initial condition, and quantities such as the Loschmidt echo cannot sustain chaotic sensitivity at late times. Nevertheless, prior to reaching the steady state, the evolution may display intricate transient behavior, including significant operator growth, enhanced sensitivity to perturbations, and complex relaxation pathways. Moreover, even as $\hat{\rho}(t)\rightarrow\hat{\rho}_\mathrm{ss}$, single quantum trajectories can exhibit chaotic asymptotic dynamics~\cite{ferrari2025dissipative,mondal2025transient}, and the steady state itself may differ qualitatively between systems believed to be chaotic or regular~\cite{rufo2025quantum,richter2025integrability, prosen2013eigenvaluestatistics}.

Motivated by the correspondence between classical chaos and quantum spectral statistics, Grobe, Haake, and Sommers (GHS) conjectured that dissipative quantum systems whose classical counterparts are chaotic should exhibit spectral level repulsion~\cite{grobe1988quantum}. In this context, the relevant spectrum is that of the Lindbladian generator, which is generally distributed across the complex plane, with universal correlations governed by non-Hermitian random matrix theory rather than by Wigner--Dyson ensembles. In recent years, Lindbladian level statistics has become a standard diagnostic tool for distinguishing regular from chaotic behavior in open quantum systems; see, for example,~\cite{akemann2019universal,denisov2019universal,sa2020complex,li2021spectral,lange2021random,rubio2022integrability,sa2023signatures,gupta2024quantum,richter2025integrability,pawar2025comparative, hamazaki2018nonhermitianmanybody, hamazaki2022lindbladianmany,feng2024boundary}.

However, this viewpoint has recently been scrutinized~\cite{villasenor2024breakdown}. Because generic Lindbladian evolution leads to a unique steady state, level repulsion in the Lindbladian spectrum cannot signal persistent late-time chaos in the same sense as in closed Hamiltonian systems. However, this observation was refined in~\cite{mondal2025transient}, which suggested that Lindbladian level statistics primarily encode transient chaotic behavior. In particular, they reflect the structure of intermediate-time dynamics, where nontrivial amplification, sensitivity to perturbations, and complexity growth may arise before eventual relaxation. The distinction between asymptotic and transient chaotic behavior therefore indicates that Lindbladian spectral correlations should not be interpreted as evidence of sustained chaotic motion.

In this work, we show that Lindbladian spectral correlations cannot generically be used as a diagnostic for transient chaotic behavior.
As a result, the GHS conjecture does not hold, even at intermediate times. 
The properties of Lindbladian and Hamiltonian spectra are conceptually different in both their dynamical interpretation and their sensitivity to perturbations.
For Hamiltonians, only energy differences enter the dynamics, and even small changes in these differences can have appreciable effects at sufficiently late times because the corresponding modes do not decay.
Lindbladian modes instead have both an oscillation frequency and a decay rate, set by the imaginary and negative real parts of their eigenvalues.
A perturbation of a rapidly decaying mode must therefore be strong enough to have an appreciable effect before that mode decays; otherwise, changes to the mode and its eigenvalue may remain dynamically negligible at all times.
Level statistics, however, probe much finer spectral scales, leaving room for weaker perturbations that change the statistics without appreciably affecting the dynamics.

This distinction is strengthened by non-normality: Hamiltonians are normal, so small perturbations induce small spectral changes, whereas Lindbladians are typically non-normal and can exhibit extreme spectral sensitivity~\cite{trefethen2020spectra}.
We show that such strong non-normality can be linked to a Liouvillian skin effect~\cite{haga2021liouvillian,ehrhardt2024exploring}.

We illustrate these findings using two simple examples: a driven quantum harmonic oscillator in contact with a thermal bath and a local lattice model based on the Hatano--Nelson construction.
For the oscillator, the exact spectrum is regular, while even tiny numerical errors are sufficient to produce full-blown level repulsion despite the completely regular dynamics.
For the lattice model, we explicitly add a perturbation that vanishes with system size and changes the expectation value of any observable by only a tiny amount at all times, even for small systems.
Without this perturbation, the spectral statistics are Poisson; with it, they are Ginibre.

\section{Results}
\subsection{Lindbladian level-repulsion}
\begin{figure}[!ht]
\includegraphics[width=\textwidth]{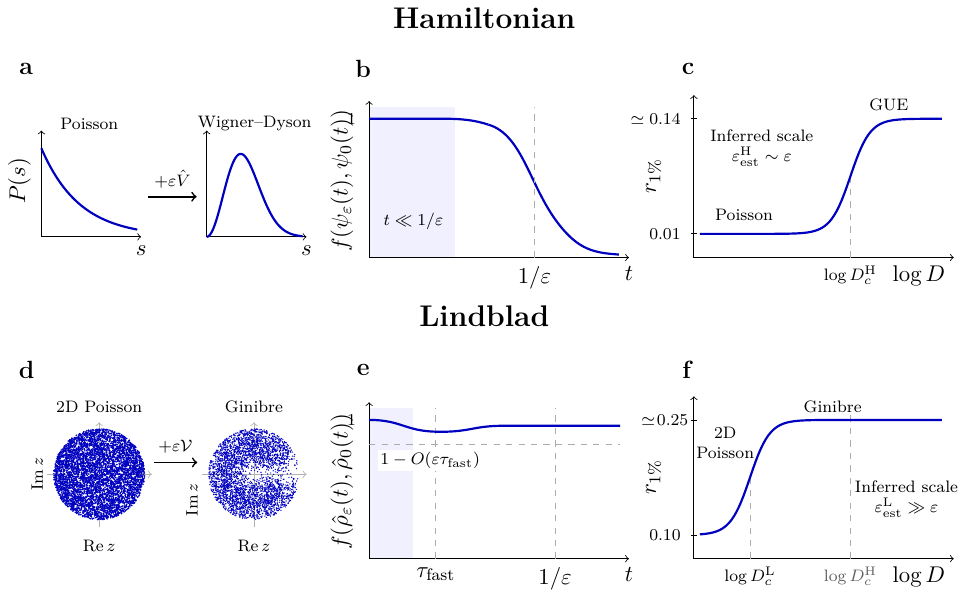}
\caption{%
\textbf{Spectral sensitivity versus dynamical stability.}
(a) Hamiltonian level statistics: a weak generic perturbation changes the normalized Hamiltonian level
statistics from Poisson to Wigner--Dyson. 
(b) Hamiltonian fidelity: the fidelity between the
perturbed and unperturbed Hamiltonian evolutions nevertheless remains close
to unity for $t\ll1/\varepsilon$. 
(c,f) We consider the crossover  in the
lower $1\%$ quantile $r_{1\%}$, defined by $\Pr(r\le r_{1\%})=0.01$, where
$r=|\lambda^{\rm NN}-\lambda|/|\lambda^{\rm NNN}-\lambda|$ as a function of Hilbert-space dimension.
(c) Hamiltonian level repulsion onset: the spectral estimate \eqref{eq:chaos-crossover-scale} defined by the onset $D_c^H$ gives a crude estimate of the strength of the term $\varepsilon\hat{V}$ that gave rise to the level-repulsion. 
(d) Lindbladian level statistics: for Lindbladians, a weak perturbation can likewise change the complex
spacing ratio distribution from the uniform 2D-Poisson disk to the
characteristic Ginibre bitten donut. 
(e) Lindbladian fidelity: when the dynamics relaxes to a
stable steady state, the fidelity decreases by at most $\mathcal O (\varepsilon \tau_{mix})$
and remains close to unity also for $t\gtrsim1/\varepsilon$.
(f) Lindbladian level repulsion: for Lindbladians the onset of level repulsion, $D_c^L$, generically says nothing about the strength of the term which caused it. 
The gray vertical line marks $D_c^{\rm H}$ from (c).
}
\label{fig:hamiltonian_crossover}
\end{figure}

The Lindblad master equation describes the most general Markovian, completely positive and trace-preserving (CPTP) continuous evolution of open quantum systems~\cite{lindblad1976generators,breuer2002theory}. 
For a system with density matrix $\hat{\rho}$ and Hamiltonian $\hat{H}$ governing the coherent part of the dynamics, the master equation takes the form
\begin{equation}
    \frac{d\hat{\rho}}{dt} = \mathcal{L}\!\left[\hat{\rho}\right] 
    \equiv -i\left[\hat{H}, \hat{\rho}\right] 
    + \sum_j\gamma_j\mathcal{D}_{\hat{L}_j}\!\left[\hat\rho\right],
    \label{eq:lindblad}
\end{equation}
with the dissipator
\begin{equation}
    \mathcal{D}_{\hat{L}_j}\!\left[\hat\rho\right]= 
        \hat{L}_j \hat{\rho} \hat{L}_j^\dagger 
        - \frac{1}{2} \left\{\hat{L}_j^\dagger \hat{L}_j, \hat{\rho}\right\},
\end{equation}
where $\hat{L}_j$ are the Lindblad jump operators with rates $\gamma_j \ge 0$. The dissipative term encodes the influence of the environment—such as dissipation, decoherence, and incoherent pumping—and therefore the state $\hat{\rho}$ generally represents a mixed state.
If diagonalizable, $\mathcal{L}$ has a spectrum $\mu_n$ with corresponding eigenoperators $\hat{o}_n$, i.e. $\mathcal{L}\left[\hat{o}_n\right]=\mu_n\hat{o}_n$. 
The CPTP property implies that, among the eigenoperators, only the steady state(s), $\mathcal{L}\left[\hat{\rho}_\mathrm{ss}\right]=0$, represents a physical density matrix, since all other eigenvectors must be traceless~\cite{minganti2019quantum,ehrhardt2024exploring}. 

Another consequence of the CPTP structure is that Lindblad time evolution is stable under small perturbations. 
Consider two Lindbladians related by a perturbation with operator trace norm\footnote{
We write $\|\hat{X}\|_p$ for the Schatten $p$-norm of an operator. For $p=1$, this is the trace norm,
$\|\hat{X}\|_1=\text{Tr}\sqrt{\hat{X}^\dagger\hat{X}}$, and for $p=2$ it is the Frobenius norm. 
The superscript $\mathrm{op}$ distinguishes the operator norm induced from the Schatten $p$-norm itself.
The operator (or induced) Schatten norm is defined as
$    \|\mathcal{O}\|^{\mathrm{op}}_p
    =
    \sup_{\hat{X}\neq0}
    \frac{\|\mathcal{O}[\hat{X}]\|_p}{\|\hat{X}\|_p} $.} 
$\varepsilon>0$,
\begin{align}
    \mathcal{L}_\varepsilon=\mathcal{L}+\varepsilon\mathcal{V},\qquad 
    \|\mathcal{V}\|^{\mathrm{op}}_1=1,
\end{align}
and assume that both $\mathcal{L}$ and $\mathcal{L}_\varepsilon$ generate CPTP evolutions. The exact variation-of-parameters formula~\cite{pazy2012semigroups} (also called Duhamel's identity) gives
\begin{equation}
    e^{t\mathcal{L}_\varepsilon}-e^{t\mathcal{L}}
    =
    \varepsilon\int_0^t
    e^{(t-s)\mathcal{L}_\varepsilon}
    \mathcal{V}
    e^{s\mathcal{L}}\,ds.
\end{equation}
Denoting $\hat{\rho}_\varepsilon(t)=e^{t\mathcal{L}_\varepsilon}[\hat{\rho}_0]$ and $\hat{\rho}(t)=e^{t\mathcal{L}}[\hat{\rho}_0]$, and using $\|\hat{\rho}(t)\|_1=1$ (which follows from $\hat \rho$ being a density matrix) together with the fact that both $\mathcal V$ and $e^{(t-s)\mathcal{L}_\varepsilon}$ are contractive with respect to the trace norm, we obtain
\begin{align}
     \|\hat{\rho}_\varepsilon(t)-\hat{\rho}(t)\|_1
    \leq \varepsilon
    \int_0^t \| e^{(t-s)\mathcal{L}_\varepsilon}\mathcal{V} e^{s\mathcal{L}}[\hat{\rho}_0]\|_1 ds
    \leq
    \varepsilon t.
    \label{eq:trace-distance-inequality}
\end{align}
Thus, a perturbation of strength $\varepsilon$ changes the physical state by at most $\mathcal O(\varepsilon t)$, and an order-one change cannot occur before times $\sim1/\varepsilon$.
To express the same stability in terms of the overlap between the physical states, rather than their norm distance, we also consider the Uhlmann fidelity $F$ and its square root $f$~\cite{nielsen2010quantum},
\begin{align}\label{ufid}
    \sqrt{F(\hat{\rho},\hat{\sigma})}
    =
    f(\hat{\rho},\hat{\sigma})
    =
    \operatorname{Tr}
    \sqrt{
    \sqrt{\hat{\rho}}\,
    \hat{\sigma}\,
    \sqrt{\hat{\rho}}
    }.
\end{align}
The Uhlmann fidelity is related to the trace distance through the inequality %
$
    1-f(\hat{\rho},\hat{\sigma})
    \leq
    \frac{1}{2}\|\hat{\rho}-\hat{\sigma}\|_1
$%
~\cite{fuchs1999cryptographic}, which together with \eqref{eq:trace-distance-inequality} implies
\begin{align}
    1-f\!\left(\hat{\rho}_\varepsilon(t),\hat{\rho}(t)\right)
    \leq
    \varepsilon t/2.
    \label{eq:fidelity}
\end{align}
Thus, also when measured in terms of fidelity, changes of order one cannot occur before times $\sim1/\varepsilon$.

We now contrast this stability of the time evolution with the much greater sensitivity of unfolded level statistics. 
Before doing so, let us first briefly review the situation in closed quantum systems.
Hamiltonian evolution is a special case of Lindblad evolution and is therefore stable in precisely the sense discussed above. 
Level repulsion, by contrast, concerns spectral structure on the scale of the local mean level spacing~\cite{haake1991quantum}. 
Given an ordered spectrum
$\{E_n\}$ with $E_{n+1}>E_n$, after resolving symmetries and removing exact degeneracies, we define the bare spacings
$\delta_n=E_{n+1}-E_n$. 
In the unfolded spectrum, the smooth component of the density of states is removed and the spacings are rescaled according to
$s_n=\delta_n/\Delta(E_n)$, where $\Delta(E)$ is the local mean level spacing, such that $\langle s\rangle=1$. 
We also disregard the spectral edges and retain only a finite fraction of eigenvalues in the bulk. 
Level repulsion then refers to the suppression of small unfolded spacings, namely to a spacing distribution satisfying $P(s)\rightarrow0$ as $s\rightarrow0$, i.e. the eigenvalues are correlated.

For random matrices drawn from the Gaussian ensembles—i.e., with probability density scaling as $e^{-\mathrm{Tr}(\hat{H}^2/\sigma^2)}$, where $\sigma$ sets an energy scale---the distribution $P(s)$ of unfolded level spacings follows the Wigner–Dyson distributions~\cite{wigner1955characteristic,dyson1962threefold,haake1991quantum,mehta2004random}. These take the form $P(s)\sim s^\beta$ for small spacings $s$, where the exponent $\beta$ depends on the time-reversal symmetry class: broken time-reversal symmetry ($\beta=2$), time-reversal symmetry squaring to $+1$ ($\beta=1$), or time-reversal symmetry squaring to $-1$ ($\beta=4$)~\cite{dyson1962threefold}. In the limit of infinite matrix dimension, the set random of matrices whose unfolded bulk level spacings deviate from Wigner–Dyson statistics has measure zero~\cite{erdos2011universality,erdos2012bulk}.

The central idea is that physical Hamiltonians in the thermodynamic limit exhibit the same universal level statistics~\cite{haake1991quantum}, once symmetries are properly accounted for. Thus, in the absence of fine-tuning---by restricting to individual symmetry sectors---Hamiltonians of sufficiently large dimension are expected to display Wigner–Dyson statistics and the associated level repulsion determined by $\beta$. This behavior is commonly associated with quantum chaos in the late-time dynamics. 
In contrast, according to the Berry–Tabor conjecture~\cite{berry1977level}, systems with a classically integrable limit are expected to have uncorrelated nearby eigenvalues and, if these eigenvalues are random, Poisson-distributed level spacings, $P(s)=e^{-s}$.

A generic integrability-breaking perturbation produces a crossover to Wigner--Dyson statistics when its matrix elements become comparable to the local mean level spacing. 
Since generically the band-width grows slower than Hilbert-space dimension, the spectrum becomes increasingly dense with system size, and this crossover scale commonly tends to zero in the thermodynamic limit. 
Thus, for any fixed generic perturbation $\varepsilon>0$, a sufficiently large system generically displays Wigner--Dyson statistics even though the perturbation is arbitrarily small on microscopic energy scales. 

After unfolding, the distinction between having Wigner--Dyson statistics or not is therefore an on-or-off question: the unfolding procedure removes the absolute energy scale that controls the crossover. 
Importantly, there is no contradiction between this spectral sensitivity and the perturbative stability of the time evolution. 
Wigner--Dyson statistics are associated with dynamical signatures of quantum chaos such as eigenstate thermalization, operator and out-of-time-order-correlator growth, and increasing circuit complexity~\cite{deutsch1991quantum,srednicki1994chaos,maldacena2016bound,brown2018second}. 
Nevertheless, the bound derived above implies that an $\varepsilon$-dependent change in the dynamics cannot become order one before times of order $1/\varepsilon$. 
The fidelity between the perturbed and unperturbed evolutions therefore remains close to unity for $t\ll1/\varepsilon$, even when their unfolded level statistics are already Wigner--Dyson and  Poisson, respectively.
Whether and how rapidly the fidelity subsequently decays is model dependent. 
This distinction is illustrated schematically in Fig.~\ref{fig:hamiltonian_crossover}(a,b).
%
%

Analogously to the Hermitian case, for open quantum systems where the spectrum is in general complex, level repulsion is also expected to signal chaotic dynamics.
This idea was further developed in~\cite{sa2020complex}, where the complex spacing ratio
\begin{equation}\label{csr}
    z_n = \frac{\mu_n^{\mathrm{NN}} - \mu_n}{\mu_n^{\mathrm{NNN}} - \mu_n},
\end{equation}
was introduced, with $\mu_n^{\mathrm{NN}}$ and $\mu_n^{\mathrm{NNN}}$ denoting the nearest and next-nearest neighbors, respectively. Level repulsion manifests itself in both the radial and angular components of $z_n$, producing the characteristic “bitten-donut’’ distribution $P(z)$, whereas non-chaotic evolution leads to a distribution that approximately fills the unit disk without strong correlations. By now, level statistics has become a widely used diagnostic for distinguishing regular from chaotic behavior in dissipative systems; see, for example,~\cite{akemann2019universal,denisov2019universal,sa2020complex,li2021spectral,lange2021random,rubio2022integrability,sa2023signatures,gupta2024quantum,richter2025integrability,pawar2025comparative, hamazaki2020universality}.

The Lindblad setting, however, allows for a sharper separation
between spectral statistics and physical dynamics.
Consider a finite-dimensional Lindbladian with a unique attracting
steady state and a finite mixing time $\tau_{\mathrm{mix}}$.
More precisely, suppose that the evolution of any Hermitian
traceless operator $\hat{X}$ satisfies
\begin{equation}
    \|
    e^{t\mathcal{L}_\varepsilon}[\hat{X}]
    \|_1
    \leq
    2 e^{-t/\tau_{\mathrm{mix}}}
    \|\hat{X}\|_1.
\end{equation}
Combining this mixing bound with the stability of the time
evolution discussed above, we find that the fidelity satisfies
\begin{equation}
    1-f\!\left(
    \hat{\rho}_\varepsilon(t),\hat{\rho}(t)
    \right)
    \leq
    \varepsilon
    \min\{
    t/2,\tau_{\mathrm{mix}}
    \} .
    \label{eq:finite-change}
\end{equation}
Thus, when $\varepsilon \tau_{\mathrm{mix}}\ll1$, the perturbed
and unperturbed evolutions remain close at all times, rather
than only for $t\ll1/\varepsilon$.
Before the mixing time this follows from perturbative stability,
while after the mixing time both systems have relaxed to nearby
steady states and essentially no further dynamics occurs.
Nevertheless, an arbitrarily small perturbation may already
have changed the bulk level statistics from uncorrelated to
strongly level-repelling.

Moreover, the time scale relevant to this argument need not be
the full mixing time of the system.
The slowly decaying modes lie near the spectral edge, whereas
the bulk statistics concern more rapidly decaying modes.
One can therefore imagine constructing a perturbation that
changes the bulk statistics while leaving the slowly decaying
modes unchanged.
Its effect on the physical dynamics can then decay on a much
shorter time scale, even while the system as a whole continues
to exhibit slow dynamics.
This time scale can remain short even when the full mixing time
diverges with system size or is infinite.
Thus, if the perturbation needed to change the bulk statistics
becomes arbitrarily weak as the Hilbert-space dimension $D$
increases, two models can have qualitatively different spectral
statistics yet arbitrarily small differences in the expectation
values of all observables of bounded norm, uniformly in time.
This separation is illustrated schematically in
Fig.~\ref{fig:hamiltonian_crossover}(d,e): the Lindbladian level
statistics change under the perturbation $\varepsilon\mathcal{V}$,
whereas the fidelity between the corresponding physical states
remains close to unity for arbitrarily long times.

The distinction from the Hamiltonian case is even sharper when
considering the onset of chaos.
In the Hamiltonian case if we consider $\hat{H}_\varepsilon=\hat{H}_0+\varepsilon\hat{V}$
with $\hat{H}_0$ an integrable reference model, we can estimate $\varepsilon$ by considering the onset of level repulsion.
In Fig.~\ref{fig:hamiltonian_crossover}(c) we track the suppression of close pairs using the lower $1\%$ quantile
$r_{1\%}$, defined by $\Pr(r\le r_{1\%})=0.01$, with
$r=|\lambda^{\rm NN}-\lambda|/|\lambda^{\rm NNN}-\lambda|$,
where NN and NNN denote the nearest and second-nearest eigenvalues.
Let $\Delta_{\rm H}(D)$ denote the typical spacing between neighboring
Hamiltonian eigenvalues in the bulk window used to compute the
statistics, i.e., the inverse local density of levels.
For example, using the bandwidth $W(D)\sim J\log D$ of a local
many-body Hamiltonian gives the crude estimate
$\Delta_{\rm H}(D)\sim W(D)/D\sim J\log D/D$,
where $J$ is a typical local energy scale.
The crossover occurs when the perturbation appreciably mixes nearby
levels, so the spacing at $D_c^{\rm H}$ gives a crude estimate of the
scale of the non-integrable term
\begin{equation}
    \varepsilon_{\rm est}^{\rm H}
    \sim\Delta_{\rm H}(D_c^{\rm H})
    \sim\frac{W(D_c^{\rm H})}{D_c^{\rm H}}.
    \label{eq:chaos-crossover-scale}
\end{equation}
With $\hbar=1$, the inverse of this estimate gives a corresponding
time scale,
\begin{equation}
    \tau_{\rm est}^{\rm H}
    \equiv\frac{1}{\varepsilon_{\rm est}^{\rm H}}
    \sim\frac{D_c^{\rm H}}{W(D_c^{\rm H})}.
    \label{eq:chaos-crossover-time}
\end{equation}
This provides a rough estimate of the time scale on which
integrability breaking becomes dynamically important.

When the perturbation is known explicitly, its strength can of course
be used directly; the point is that the onset of level repulsion as
the system size increases also provides an indirect estimate from spectral information alone.
For Hamiltonians, the key constraint underlying this estimate is
Weyl's inequality: for $\|\hat{V}\|_\infty=1$, every eigenvalue shifts by at most $\varepsilon$.
(The operator norm $\|\hat{V}\|_\infty$ controls the induced trace norm of the
Hamiltonian commutator through $\|-i[\hat{V},\cdot]\|^{\mathrm{op}}_1\leq2\|\hat{V}\|_\infty$,
and therefore sets the perturbation scale in the trace-distance bound
\eqref{eq:trace-distance-inequality} for Hamiltonian dynamics.)
A perturbation therefore cannot rearrange eigenvalues over energy
scales parametrically larger than its strength.
In the Lindbladian case no analogous bound applies.
The crossover can occur while the typical separation
$\Delta_{\rm L}(D_c^{\rm L})$ of neighboring bulk eigenvalues is still
much larger than $\varepsilon$.
The onset of level repulsion in a Lindbladian therefore does generically not
reflect a physical time scale at all.
This contrast is illustrated schematically in
Fig.~\ref{fig:hamiltonian_crossover}(c,f).

Let us now turn to a first example that illustrates the problem of drawing any conclusions from the bulk spectra of a Lindbladian.
We consider a driven quantum harmonic oscillator subjected to incoherent loss and gain, i.e. coupled to a thermal bath,
\begin{equation}\label{dho}
\hat H = \omega \hat a^\dagger \hat a + \eta (\hat a + \hat a^\dagger),\qquad
{\mathcal{D}}[\hat\rho] = \gamma_1 \mathcal{D}_{\hat L_1}[\hat\rho] + \gamma_2 \mathcal{D}_{\hat L_2}[\hat\rho].
\end{equation}
Here, $\hat L_1=\hat a$ and $\hat L_2=\hat a^\dagger$. This model is quadratic, and therefore any initial Gaussian state remains Gaussian under the dynamics. Expectation values $A(t) = \langle \hat A \rangle$ obey the corresponding classical equations of motion, which can often be obtained in closed analytical form~\cite{SuppMat}.  
For $\gamma=\gamma_1-\gamma_2>0$, the Lindbladian spectrum can be obtained using the method of third quantization~\cite{prosen_third_2008}, generalized to bosonic models~\cite{prosen_quantization_2010,kim_third_2023}, yielding
\begin{equation}
    \mu_{n,m} = -\frac{\gamma}{2} (n+m) + i\,\omega (n-m),\qquad n,m \in \mathbb{N}_0.
    \label{eq:hospec}
\end{equation}
This expression extends the equidistant harmonic oscillator spectrum into the complex plane. 

\begin{figure}[!ht]
    \includegraphics[width=\textwidth]{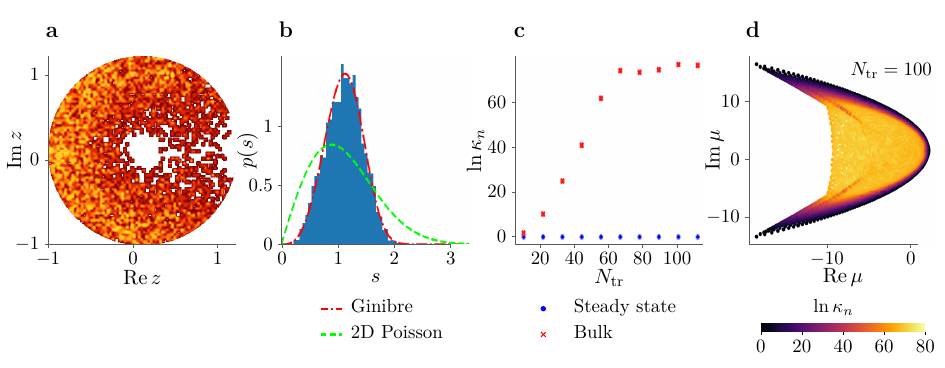}
\caption{%
\textbf{Spectral statistics and eigenvalue condition numbers of the driven harmonic oscillator.}
(a) Two-dimensional density of the complex spacing ratio~(\ref{csr}) for the truncated driven harmonic oscillator~(\ref{dho}).
The inset shows the analytical Ginibre density.
White bins within the unit disk, including those near the centre, contain no samples because of limited statistics.
The averages are $\langle r\rangle\approx0.7483$ and $-\langle\cos\theta\rangle\approx0.3013$, compared with the Ginibre predictions $\langle r\rangle\approx0.73810$ and $-\langle\cos\theta\rangle\approx0.24051$.
Their convergence toward the Ginibre predictions with increasing truncation is shown in the Supplementary Material~\cite{SuppMat}.
(b) Histogram of unfolded nearest-neighbour spacings, compared with the two-dimensional Poisson distribution (green dashed curve) and the Ginibre prediction (red dash-dotted curve).
Panels (a,b) use $N_{\mathrm{tr}}=135$, with their bulk selection and the unfolding procedure for (b) described in the Methods section.
(c) The eigenvalue condition numbers~(\ref{eq:def-cond}) for the steady-state eigenvalue (blue circles) and a representative bulk eigenvalue (red crosses), as a function of the truncation $N_{\mathrm{tr}}$.
The steady-state condition number remains approximately constant, whereas the bulk condition number grows exponentially.
The knee near $N_{\mathrm{tr}}\approx60$ reflects the numerical precision limit for resolving exponentially small left--right eigenvector overlaps.
(d) Numerically obtained spectrum $\mu_n$ at $N_{\mathrm{tr}}=100$, colored by $\ln\kappa_n$.
Bulk eigenvalues have much larger condition numbers than those near the steady state. In the Supplementary Material~\cite{SuppMat} we analytically show that in the thermodynamic limit (i.e. the untruncated model), all Petermann factors, except for the steady state, diverge.
The parameters are $\omega=0$, $\eta=\sqrt{2}/4$, $\gamma_2=0.05$, and $\gamma_1=0.05005$.
}
    \label{fig:levelstat}
\end{figure}

We numerically diagonalize the Lindbladian and analyze its spectral statistics, namely the spacing distribution $P(s)$ and the complex spacing ratio~(\ref{csr}). More precisely, when diagonalizing the oscillator model we introduce a bosonic truncation cutoff $N_\mathrm{tr}$.
 In Fig.~\ref{fig:levelstat}(a-b) we show that the spectra display clear level repulsion consistent with the cubic Ginibre behavior~\cite{ginibre1965statistical}, and that the complex spacing ratio forms the familiar “bitten-donut’’ pattern widely interpreted as a hallmark of chaotic non-Hermitian dynamics~\cite{sa2020complex}. 
This is in apparent contradiction with the fact that the open driven harmonic oscillator is integrable. To understand how this comes about we need to first develop a better understanding of how perturbations affect the spectrum of non-normal operators.

\subsection{Spectral instability and non-normal operators}
Comparing the analytical spectrum~(\ref{eq:hospec}) with the numerical results in Fig.~\ref{fig:levelstat} reveals a striking discrepancy: the regular oscillator spectrum is replaced by a spectrum with pronounced level repulsion.
To understand its origin, we examine the non-normality of the Lindbladian and the resulting sensitivity of its eigenvalues to perturbations.

A matrix $\hat{M}\in\mathbb{C}^{l\times l}$ is normal if it commutes with its Hermitian conjugate, $[\hat{M},\hat{M}^\dagger]=0$.
Otherwise, it is non-normal and therefore also non-Hermitian, although a non-Hermitian matrix need not be non-normal.
If $\hat{M}$ is diagonalizable, its right and left eigenvectors satisfy
\begin{align}
    \hat{M}\ket{\xi_n}
    &=\lambda_n\ket{\xi_n},
    &
    \hat{M}^\dagger\ket{\zeta_n}
    &=\lambda_n^*\ket{\zeta_n}.
\end{align}
For a non-normal matrix, this basis cannot be chosen orthonormal, but it can be normalized biorthogonally~\cite{golub2013matrix,moiseyev2011non},
\begin{align}
    \braket{\zeta_m|\xi_n}
    &=\delta_{mn},
    &
    \hat{\mathbb{I}}
    &=\sum_{n=1}^{l}\ketbra{\xi_n}{\zeta_n}.
\end{align}

To quantify the resulting spectral sensitivity, consider perturbing $\hat M^{(0)}$ by $\varepsilon\hat M^{(1)}$,
\begin{equation}
    \hat M(\varepsilon)
    =\hat M^{(0)}+\varepsilon\hat M^{(1)}.
\end{equation}
The eigenvalue condition number gives the largest first-order shift per unit perturbation strength
\begin{equation}
    \max_{\|\hat M^{(1)}\|^{\mathrm{op}}_p=1}
    \left|
    \lambda_n(\varepsilon)-\lambda_n^{(0)}
    \right|
    =\varepsilon\kappa_p(\lambda_n^{(0)})+O(\varepsilon^2).
    \label{eq:def-cond}
\end{equation}
First-order perturbation theory gives
\begin{equation}
    \lambda_n^{(1)}
    =\frac{
        \bra{\zeta_n^{(0)}}\hat M^{(1)}\ket{\xi_n^{(0)}}
    }{
        \braket{\zeta_n^{(0)}|\xi_n^{(0)}}
    }.
    \label{eq:perturbation-theory}
\end{equation}
Maximizing over perturbations of unit norm yields\footnote{H\"older's inequality gives $|\bra{\zeta}\hat M^{(1)}\ket{\xi}|\leq\|\ket{\zeta}\|_q\|\hat M^{(1)}\|^{\mathrm{op}}_p\|\ket{\xi}\|_p$.
The bound is attained by a rank-one matrix $\hat M^{(1)}_\star=\ketbra{A}{B}$, where $\|\ket A\|_p=\|\ket B\|_q=1$, $|\braket{\zeta|A}|=\|\ket\zeta\|_q$, and $|\braket{B|\xi}|=\|\ket\xi\|_p$.}
\begin{equation}
    \kappa_p(\lambda_n^{(0)})
    =\frac{
        \|\ket{\zeta_n^{(0)}}\|_q
        \|\ket{\xi_n^{(0)}}\|_p
    }{
        \left|\braket{\zeta_n^{(0)}|\xi_n^{(0)}}\right|
    }.
\end{equation}
Here, $1/p+1/q=1$, including $q=\infty$ for $p=1$.
This expression is independent of the normalization of the eigenvectors.
For $p=2$, if both eigenvectors have unit norm, the condition number is simply the inverse magnitude of their overlap.
For normal matrices, the left and right eigenvectors can be chosen identical and orthonormal, giving $\kappa_2=1$.
For non-normal matrices, unit-normalized left and right eigenvectors can have very small overlaps, producing large condition numbers and correspondingly amplified eigenvalue shifts.

For perturbations of a Lindbladian, the trace norm is physically relevant because it controls changes in expectation values of bounded observables and enters the stability bound~\eqref{eq:trace-distance-inequality}.
Here, however, we are concerned with the coarse scaling of the condition numbers, particularly whether they grow exponentially with the Hilbert-space dimension $D$.
Since $\|\hat{X}\|_2\leq\|\hat{X}\|_1\leq\sqrt{D}\|\hat{X}\|_2$, the superoperator norms induced by the trace and Hilbert--Schmidt norms, and the corresponding eigenvalue condition numbers, differ by at most polynomial factors in $D$.
These factors cannot change exponential growth in $D$.
We therefore use the Hilbert--Schmidt norm and its induced superoperator norm below, as they are simpler to handle analytically and numerically.


Lindbladians need not all exhibit the strong non-normality considered here.
Standard random-matrix ensembles provide a useful contrast: in the complex Ginibre ensemble, typical bulk eigenvalue condition numbers grow polynomially with matrix dimension~\cite{chalker1998eigenvector,trefethen2020spectra}.
The exponential sensitivity in our examples therefore involves additional structure in the generator that is not specified by its level statistics.
Agreement with Ginibre eigenvalue correlations does not by itself imply comparable eigenvector properties or spectral stability.

A physical mechanism that can generate strong non-normality is nonreciprocal motion in the presence of a boundary, arising, for example, from asymmetric hopping or pumping.
Such motion can produce a non-Hermitian skin effect, in which an extensive set of eigenmodes accumulates near a boundary rather than being localized by disorder~\cite{okuma2020topological,zhang2022review}.
This accumulation can make the eigenvector basis strongly ill-conditioned: the matrix $\hat{\mathbf{R}}$ formed from normalized right eigenvectors develops very small singular values, so its columns become nearly linearly dependent at finite numerical precision.
The resulting sensitivity has been connected to numerical instabilities in the Hatano--Nelson model~\cite{feng2025numerical}, while boundary-sensitive spectra and relaxation in Lindbladians were studied in Ref.~\cite{feng2024boundary}.

In the oscillator model, the truncation $N_{\mathrm{tr}}$ introduces an artificial boundary in Fock space by excluding states of higher occupation.
The coherent drive and the loss and gain processes, $\mathcal D_{\hat L_1}$ and $\mathcal D_{\hat L_2}$, connect different occupations, with the imbalance between loss and gain producing nonreciprocal couplings in the associated Liouville--Fock lattice~\cite{naves2025liouville}.
The resulting boundary sensitivity concerns the Lindbladian eigenmodes and need not imply that the physical state approaches the cutoff.
Increasing the cutoff can therefore leave low-lying modes well resolved while the spectral bulk, growing in proportion to the number of low lying modes, remains strongly affected by the boundary.
A related mechanism underlies the Hatano--Nelson construction introduced below~(\ref{eq:HNmodel}).
Further details on the skin effect in Liouville space are given in the Supplementary Material~\cite{SuppMat}.

We now return to the numerical spectrum in Fig.~\ref{fig:levelstat}(d).
For $\omega=0$, the analytical spectrum~(\ref{eq:hospec}) is purely real, with $\mu_n=-n\gamma/2$ and $n\in\mathbb{N}_0$, whereas the numerical eigenvalues spread into the complex plane and display pronounced level repulsion.
The color map shows the corresponding eigenvalue condition numbers~\eqref{eq:def-cond}: eigenvalues near the steady state have relatively small condition numbers, while those in the bulk have extremely large values.
These condition numbers are so large that the discrepancy between the analytical and numerical spectra arises from numerical round-off errors (of order $10^{-16}$), here amplified by this strong non-normality.

Figure~\ref{fig:levelstat}\textbf{c} shows the natural logarithm of the condition numbers as a function of $N_{\mathrm{tr}}$, for the steady-state eigenvalue (blue circles) and a representative bulk eigenvalue (red crosses).
The steady-state condition number remains essentially constant, whereas the bulk condition number grows exponentially up to $N_{\mathrm{tr}}\approx60$ and then saturates.
This saturation reflects the finite precision available to resolve exponentially small left--right eigenvector overlaps; increasing the numerical precision extends the regime of exponential growth to larger truncations.
In the Supplementary Material~\cite{SuppMat}, we further show that the Petermann factors for each eigenvector, apart from the steady state, diverge in the thermodynamic limit, i.e. for the full untruncated model. Thus, increasing $N_{\mathrm{tr}}$ does not improve the balance between well-conditioned and ill-conditioned eigenvalues: the bulk with large condition numbers becomes increasingly dominant.

The exponential growth of the bulk condition numbers makes the spectrum sensitive to perturbations on an exponentially decreasing scale.
Already at the truncations considered here, numerical round-off is sufficient to produce Ginibre-like level repulsion.
As $N_{\mathrm{tr}}$ increases, ever smaller matrix perturbations can therefore produce appreciable changes in the bulk spectrum.
Two generator matrices can thus be arbitrarily close in norm while exhibiting qualitatively different bulk spectral statistics, obstructing a direct physical interpretation of those statistics. This will be explicitly demonstrated in the next section by introducing an analytically tractable lattice model.

\subsection{A model with both Poisson and Ginibre level statistics}
\label{subsec:localspinlattice}
We have discussed how, in the case of strong non-normality, extremely small perturbations can change the spectrum qualitatively.
Machine-precision errors alone made an integrable model exhibit level repulsion---the model including perturbations of machine-precision size \textit{has} level repulsion.
As we discussed, we can take these perturbations to decay so fast with system size that their effect on any observable vanishes at fixed times, while still changing the spectrum qualitatively.

A possible loophole preserving the physical importance of the spectrum could be that small perturbations can indeed have this effect, but physically relevant perturbations might not.
Contrary to numerical round-off errors, such perturbations must preserve CPTP properties and satisfy physical constraints, for example involving only finite-degree polynomials in bosonic creation and annihilation operators in a bosonic model, or being local in a lattice model.
A second loophole is that, if the mixing time of the model is infinite, even a very small change to the Lindbladian could eventually produce an appreciable change in observables.
A third objection (weaker than the others) is that we have not actually shown an example that goes from 2D Poisson to Ginibre statistics.
We have shown that a model with a simple, exactly solvable spectrum, and clearly regular dynamics, can acquire Ginibre statistics under an arbitrarily small perturbation, but could there be some principle forbidding a direct transition from 2D Poisson to Ginibre statistics?

In this section, we resolve these loopholes by considering a local lattice model with infinite mixing time in the thermodynamic limit and an explicit perturbation that we show numerically changes the spectral statistics from 2D Poisson to Ginibre.
At the same time, we obtain an analytic bound, uniform in $t$ and $L$, on how much the time-evolved density matrices $\hat{\rho}_\varepsilon(t)=e^{t\mathcal L_\varepsilon}[\hat{\rho}_{\mathrm{in}}]$ and $\hat{\rho}_0(t)=e^{t\mathcal L_0}[\hat{\rho}_{\mathrm{in}}]$, evolved from the same arbitrary initial state with the perturbation turned on and off, can differ:
\begin{align}
    F\bigl(\hat{\rho}_\varepsilon(t),\hat{\rho}_0(t)\bigr)
    &\geq
    \left(1-\frac{\varepsilon}{4\Gamma}\right)^2.
    \label{eq:main-bound}
\end{align}
Here, $\varepsilon$ is the perturbation strength and $\Gamma>0$ is a free parameter in the $\varepsilon=0$ Lindbladian; we set $\Gamma=1$ in all numerics.
We also choose the model such that the perturbation strength required to go from 2D Poisson to Ginibre statistics is expected to decay exponentially with $L$.
We can therefore let the perturbation decay with $L$ while retaining the change in spectral statistics: we take $\varepsilon=\varepsilon_0/L$, with $\varepsilon_0=2$ in the numerics.
This means that, by considering sufficiently large $L$, one can make the predictions for \textit{all} physical observables with uniformly bounded operator norm arbitrarily close between the two evolutions, independently of how long we wait.
At any given precision in these observables, there is thus an $L$ above which the two theories are indistinguishable for all times.
This example thus shows that the same physical dynamics, to any prescribed precision, can be associated with either Ginibre or 2D Poisson statistics.

A simple way to combine 2D Poisson statistics with the strong spectral sensitivity is to use the Hatano--Nelson model~\cite{hatano1996localization}, which describes a particle with asymmetric hopping on a one-dimensional lattice.
Its simplest version has hopping in only one direction and, on a ring, is defined by the non-Hermitian Hamiltonian
\begin{align}
    \hat{K}&=\sum_{i=1}^{L}\alpha_i\ket{i}\!\bra{i}
      +i\sum_{i=1}^{L}g_i\ket{i+1}\!\bra{i},
    \label{eq:HNmodel}
\end{align}
where $\ket{L+1}\equiv\ket{1}$.
We make one further simplification by setting one hopping amplitude to zero, without loss of generality $g_L=0$, and taking $g_j=g>0$ for the remaining bonds.
In the position basis, this model is bidiagonal, so its eigenvalues are simply the diagonal elements $\{\alpha_j\}$.
For sufficiently large $g$ relative to the diagonal disorder, all\footnote{Exponential growth of condition numbers associated with the Hatano--Nelson skin effect is discussed in Ref.~\cite{nakai2024topological}, but is not generally guaranteed for every eigenvalue; here, the bidiagonal structure allows us to calculate the left and right eigenvectors analytically and derive the bound $\kappa_j(\hat{K})\geq g^{L-1}/\prod_{m\neq j}|\alpha_j-\alpha_m|$. Details are given in the supplemental material~\cite{SuppMat}.} eigenvalues have condition numbers that grow exponentially with system size, so a crossover from Poisson to Ginibre statistics under a very small generic perturbation is expected.

\begin{figure}[!th]
\includegraphics[width=\textwidth]{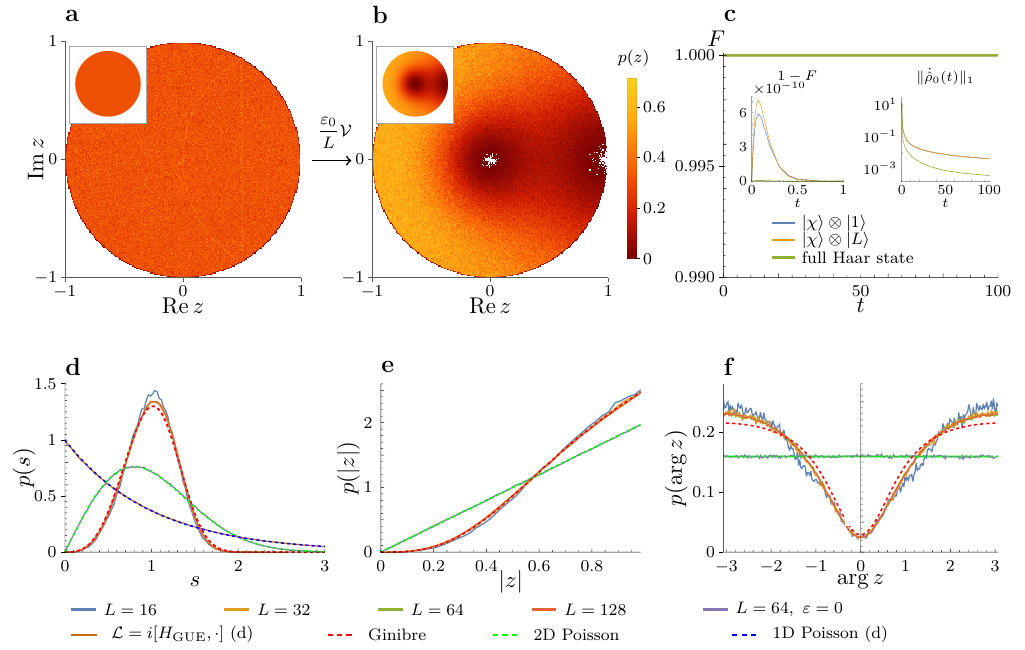}
\caption{%
\textbf{Spectral sensitivity and dynamical stability of the Hatano--Nelson construction and level statistics of a purely Hamiltonian Lindbladian.}
(a,b) Numerical complex spacing-ratio distributions for $L=64$ before ($\varepsilon  = 0 $) and after adding the perturbation $\varepsilon = (\varepsilon_0/L)\mathcal{V}$ (insets show analytical Poisson and Ginibre distributions).
The densities are estimated from two-dimensional histograms with square bins of width $\Delta(\operatorname{Re}z)=\Delta(\operatorname{Im}z)=0.01$.
(c) Fidelity $F(\hat{\rho}_\varepsilon(t),\hat{\rho}_0(t))$ for $L=128$, starting from a particle at site $1$ or $L$ with the same Haar-random spin state $\lvert\chi\rangle$, or a Haar-random pure state in the full Hilbert space.
The insets show the early-time infidelity and $\|\dot{\hat{\rho}}_0(t)\|_1$.
(d--f) Distributions of the unfolded nearest-neighbour spacing $s$, the ratio modulus $|z|$ and its argument $\arg z$, respectively.
Solid curves show probability densities estimated from normalized histograms with bin widths $\Delta s=\Delta |z|=\Delta(\arg z)=0.025$; red and green dashed curves show the Ginibre and $2D$ Poisson predictions, respectively.
The solid curves for the lattice model show spectral statistics in the $+$ sector (and also describe the $S\to\infty$ limit of the full Lindbladian spectrum in the extended model).
Panel (d) also shows the unfolded spacings for the purely Hamiltonian Lindbladian $\mathcal L=i[H_{\mathrm{GUE}},\cdot]$, with Hamiltonian dimension $D=128$, in brown; the blue dashed curve is the one-dimensional Poisson prediction $p(s)=e^{-s}$.
Disorder averaging is performed by pooling the selected bulk samples from $3000$ realizations for each system size in the $+$ sector; the GUE data use $3000$ independently drawn Hamiltonians.
The lattice-model parameters are $g=1.5$, $W=\Gamma=1$, $\lambda=4$, and $\varepsilon_0=2$ for the perturbed data.
All plotted data are reliable within the estimated numerical errors.
The numerical precision tests, unfolding procedure, and bulk selection are described in the Methods section.
}
\label{fig:HNetcmodel}
\end{figure}

This is, of course, not a Lindbladian but a non-Hermitian Hamiltonian.
However, Hatano--Nelson dynamics can be realized within a Lindbladian~\cite{mcdonald2022nonequilibrium}.
We introduce an internal spin-$1/2$ degree of freedom and a weak $U(1)$ symmetry generated by spin-$z$.
Writing, e.g., $\hat{\rho}_{\uparrow\downarrow}=\bra{\uparrow}\hat{\rho}\ket{\downarrow}$ for the position-space blocks, we decompose the density matrix as $\hat{\rho}=\hat{\rho}_{\mathrm{pol}}+(\ket{\uparrow}\!\bra{\downarrow}\otimes\hat{\rho}_{\uparrow\downarrow}+\mathrm{H.c.})$, where $\hat{\rho}_{\mathrm{pol}}=\sum_{a=\uparrow,\downarrow}\ket{a}\!\bra{a}\otimes\hat{\rho}_{aa}$.
The weak symmetry ensures that the three supercharge sectors evolve independently, i.e., $\mathcal L(\hat{\rho})=\mathcal L_{\mathrm{pol}}(\hat{\rho}_{\mathrm{pol}})+\ket{\uparrow}\!\bra{\downarrow}\otimes\mathcal L_+(\hat{\rho}_{\uparrow\downarrow})+\ket{\downarrow}\!\bra{\uparrow}\otimes\mathcal L_-(\hat{\rho}_{\downarrow\uparrow})$, and the full spectrum is therefore the union of the spectra of $\mathcal L_{\mathrm{pol}}$ and $\mathcal L_\pm$.
We study spectral statistics separately within each symmetry sector, focusing here on the two off-diagonal spin blocks $\mathcal L_\pm$.
We choose the Lindbladian such that the two off-diagonal spin blocks realize Hatano--Nelson dynamics, with the positive-charge sector given by
\begin{align}
    \mathcal L_+(\hat{\rho}_{\uparrow\downarrow})
    &=-i[\hat{K},\hat{\rho}_{\uparrow\downarrow}]_\dagger
     =-i(\hat{K}\hat{\rho}_{\uparrow\downarrow}-\hat{\rho}_{\uparrow\downarrow}\hat{K}^\dagger).
    \label{eq:HN-supercharge-blocks}
\end{align}
The negative-charge sector follows by Hermitian conjugation, $\mathcal L_-(\hat{\rho}_{\downarrow\uparrow})=[\mathcal L_+(\hat{\rho}_{\uparrow\downarrow})]^\dagger$.
The eigenvalues in either sector are thus $\mu_{jk}=-i(\alpha_j-\alpha_k^*)$.
The corresponding left and right eigenoperators are outer products of the eigenvectors of $\hat{K}$, so their Hilbert--Schmidt norms and overlaps factorize, giving $\kappa_{jk}(\mathcal L_\pm)=\kappa_j(\hat{K})\kappa_k(\hat{K})$.
The exponential growth of the condition numbers therefore carries over to both off-diagonal spin sectors.
We choose $\alpha_i\sim C+\operatorname{Unif}[0,W]+i\operatorname{Unif}[0,W]$, with $C=-i[g+W+\frac{\Gamma}{2}\coth(\lambda^{-1})]$ ($\Gamma$ is the same scale as in \eqref{eq:main-bound} and $\lambda$ is a length-scale we will discuss later).
The resulting spectral statistics, shown for $\varepsilon=0$ in Fig.~\ref{fig:HNetcmodel}(a) and by the purple line in Fig.~\ref{fig:HNetcmodel}(d-f),
 are in perfect agreement with 2D Poisson statistics.


We will take the Lindbladian to be unital and the maximally mixed state to be the unique steady state, which, together with locality, guarantees that the mixing time diverges with system-size.
To see why that is, we can consider the model in an extended context where the model is just the one-particle sector of a local model with strong particle-number conservation.
In that model we can then compare an initially localized particle with the vacuum (which is stationary because it is the only state in the zero-particle sector): the two states differ only locally, so a Lieb--Robinson bound~\cite{poulin2010lieb,nachtergaele2011lieb} requires a time proportional to $L$ before the particle can occupy the opposite half of the ring with order-one probability.
Since the maximally mixed state (in the unextended model, or equivalently in the one-particle sector) assigns order-one probability to that region, the mixing time grows at least linearly with $L$; further details are given in the Supplemental Material~\cite{SuppMat}.

It remains to complete the previously specified dynamics \eqref{eq:HN-supercharge-blocks} by choosing $\mathcal L_{pol}$ that makes the full $\mathcal L_0$ CPTP generator unital and the steady-state unique.
There are in principle several possibilities; we here take
\begin{equation}
\begin{array}{lll}\label{eq:fullLindbladian}

    \bra{\uparrow}\mathcal L_{\mathrm{pol}}(\hat{\rho}_{\mathrm{pol}})\ket{\uparrow}
    & = & \displaystyle{-i[\hat{K},\hat{\rho}_{\uparrow\uparrow}]_\dagger
      +2\sum_{i=1}^{L}r_i\braket{i|\hat{\rho}_{\downarrow\downarrow}|i}\ket{i}\!\bra{i}}
\\ \\
& & \displaystyle{+\frac{g}{2}\sum_{i=1}^{L-1}
      \hat{T}_i\hat{\rho}_{\downarrow\downarrow}\hat{T}_i
+\Gamma\sum_{i,j=1}^{L}
      \frac{\braket{j|\hat{\rho}_{\uparrow\uparrow}|j}\ket{i}\!\bra{i}}
      {e^{(|i-j|_L+|i|_L+|j|_L)/\lambda}},}

\end{array}
\end{equation}
where $\hat{T}_i=(\ket{i}-\ket{i+1})(\bra{i}-\bra{i+1})$ is the local bond operator.
The analogous equation for $\hat{\rho}_{\downarrow\downarrow}$ is obtained by interchanging $\uparrow$ and $\downarrow$.
Far from the bond between sites $1$ and $L$, the on-site rates are approximately $r_i\simeq-\operatorname{Im}\alpha_i-g$.
To this value there is a correction $-\frac{\Gamma}{2}\sum_{j=1}^{L}e^{-(|i-j|_L+|i|_L+|j|_L)/\lambda}$ correcting for the addition of the $\Gamma$ term, while the first and last sites each have an additional correction $g/2$ to compensate for the missing bond.
In conventional Lindblad form, the full $\mathcal L_0$ simply corresponds to a Hamiltonian with on-site energies $\operatorname{Re}\alpha_i$ and nearest-neighbor hopping terms $ig\hat{T}_i/2$.
There are also dissipative terms on every site and bond, defined by the jump operators $\hat{L}_i^{\mathrm{on}}=\ket{\uparrow}\!\bra{\downarrow}\otimes\ket{i}\bra{i}$ and $\hat{L}_i^{\mathrm{bond}}=\ket{\uparrow}\!\bra{\downarrow}\otimes \hat{T}_i$, which appear together with their Hermitian conjugates at rates $2r_i$ and $g/2$, respectively, alongside the spin-preserving jumps represented by the $\Gamma$-term.
The on-site rates fix the remaining imaginary parts of $\alpha_i$, while the bond dissipation combines with the Hamiltonian hopping to give the one-directional hopping of strength $g$ in $\hat{K}$.
All jump rates are nonnegative for the chosen common imaginary shift of the $\alpha_i$'s. 

This choice is unital because every jump operator appears together with its
Hermitian conjugate at the same rate, including the $\Gamma$ jumps, whose
rates are symmetric under $i\leftrightarrow j$.
For a unital Lindbladian non-uniqueness of the steady-state would require a strong symmetry: an operator commuting with the Hamiltonian and all jump operators and their adjoints~\cite{frigerio1978stationary}.
Commutation with all the spin-preserving jumps forces such an operator to be proportional to the identity within each spin sector, and the spin-flip jumps then require the proportionality constants to coincide.
Thus no nontrivial strong symmetry exists, and the maximally mixed state is the unique steady state. 

We now choose a unital perturbation $\varepsilon\mathcal V$ that only acts in the off-diagonal spin sectors.
We take it to be exponentially local and concentrated near the bond between sites $1$ and $L$, so that its norm remains bounded as the system size increases.
One cannot have a perturbation only in the charge $\pm$ sectors that generates CPTP dynamics on its own. 
This is the reason for the addition of the term proportional to $\Gamma$ in Eq.~\eqref{eq:fullLindbladian}. Together with that term, the perturbation generates CPTP dynamics for $0\leq\varepsilon\leq2\Gamma$ (further details can be found in the Supplemental Material~\cite{SuppMat}).
Its restriction to the positive-charge sector is
\begin{align}
    \mathcal V_+(\hat{\rho}_{\uparrow\downarrow})
    &=
    \sum_{\substack{i_{11},i_{12}\\i_{21},i_{22}}=1}^{L}
    \frac{
        i^{n_{\{i_{ab}\}}}
        \braket{i_{12}|\hat{\rho}_{\uparrow\downarrow}|i_{22}}
        \ket{i_{11}}\!\bra{i_{21}}
    }{
        2\coth^2\!\left(\frac{1}{2\lambda}\right)
        e^{
            \frac{1}{2\lambda}\sum_{a,b=1}^{2}
            \left(
                |i_{ab}|_L
                +|i_{ab}-i_{\bar a b}|_L
                +\frac12|i_{ab}-i_{a\bar b}|_L
            \right)
        }
    }.
    \label{eq:coherence-perturbation}
\end{align}
Here, the bars interchange $1$ and $2$ (i.e., $\bar 1=2$ and $\bar 2=1$), and we take $n_{\{i_{ab}\}}$ to be a random integer.
The negative-charge sector follows by Hermitian conjugation, $\mathcal V_-(\hat{\rho}_{\downarrow\uparrow})=[\mathcal V_+(\hat{\rho}_{\uparrow\downarrow})]^\dagger$.

With this perturbation the proof of the bound~\eqref{eq:main-bound} closely follows the derivation of Eq.~\eqref{eq:trace-distance-inequality}, but we only need to consider the off-diagonal spin block $\hat{\rho}_{\uparrow\downarrow}$.
We use $\|\mathcal V_+\|^{\mathrm{op}}_1\leq b^*/2$, where $b^*=\coth(\lambda^{-1})$.
The difference is that, instead of bounding the integrand by one, we keep the decay of the unperturbed coherences: $\|e^{s\mathcal L_+}(\hat{\rho}_{\uparrow\downarrow}(0))\|_1\leq\|e^{-is\hat{K}}\|_\infty^2\|\hat{\rho}_{\uparrow\downarrow}(0)\|_1\leq\frac12e^{-\Gamma b^*s}$.
Here, the exponential bound follows from the chosen common imaginary shift of $\hat{K}$, while positivity and normalization of the initial state imply $\|\hat{\rho}_{\uparrow\downarrow}(0)\|_1\leq1/2$.
Using the fact that the diagonal spin blocks evolve identically with and without the perturbation, together with the same relation between fidelity and trace distance used in Eq.~\eqref{eq:fidelity} then gives Eq.~\eqref{eq:main-bound} uniformly in time and independently of the initial state.
Further details are given in the Supplemental Material~\cite{SuppMat}.

Finally, we set \(W=1\), \(g=3/2\), \(\Gamma=1\), \(\lambda=4\), and \(\varepsilon=\varepsilon_0/L\), with \(\varepsilon_0=2\). Figure~\ref{fig:HNetcmodel} shows the resulting spectral statistics: the complex spacing ratio (a,b), the spacing distribution \(p(s)\) (d), the radial distribution \(p(|z|)\) (e), and the argument distribution \(p(\mathrm{arg}\,z)\) (f). We observe a clear crossover from Poisson to Ginibre statistics, even though the perturbation's effect on the dynamics becomes arbitrarily small as \(L\) increases. In particular, the fidelity shown in panel (c) remains extremely close to unity, with a deviation of order \(10^{-10}\).

The construction follows a simple prescription: realize the simplest, one-directional Hatano--Nelson model in the off-diagonal spin blocks~\eqref{eq:HN-supercharge-blocks} and complete it to a spatially local, unital Lindbladian.
Its standard Lindblad form is more involved than e.g., that of the oscillator, reflecting our choice to retain this particularly simple block dynamics and obtain the analytical bound~\eqref{eq:main-bound}.
This additional structure is not needed for the phenomenon itself: in the Supplemental Material~\cite{SuppMat}, we also study a simple one-dimensional lattice model with a more conventional Lindblad form.
It exhibits the same crossover from absence of level-repulsion to Ginibre statistics while its dynamical evolution remains essentially unchanged.

Finally we want to address an additional possible concern. 
We have closed the previous loopholes, but one could wonder if there is a possibility that this behavior is only possible within a symmetry sector and cannot occur in the full spectrum of a Lindbladian.
For spin $1/2$, adding a sufficiently large Hamiltonian term proportional to $\hat{S}_z$ separates the spectra of the three sectors without changing their internal level statistics, so the local statistics of the full spectrum become a weighted average of those of the separate sectors.
We can then extend the construction to spin $S$ by replacing each pair of spin-flip jumps with a directed cycle of transitions,
\begin{align}
    \hat{L}_{m,i}^{\mathrm{on}}
    &=\ketbra{m+1}{m}\otimes\ket{i}\!\bra{i},
    &
    \hat{L}_{m,i}^{\mathrm{bond}}
    &=\ketbra{m+1}{m}\otimes \hat{T}_i,
    \label{eq:cyclic-jumps}
\end{align}
where $m=-S,\ldots,S$, $\ket{S+1}\equiv\ket{-S}$, and the rates remain $2r_i$ and $g/2$, respectively.
Every internal state has one incoming and one outgoing jump of each type at the same rate, ensuring unitality.
The recycling terms transfer population from one internal state to the next, while the loss terms in every off-diagonal spin block remain unchanged.
For $N=2$, this cycle reproduces the original pair of spin-flip jumps.
We also copy the spin-preserving $\Gamma$ term to every internal state and use the same perturbation $\mathcal V_+$ in every block $\hat{\rho}_{mn}=\bra{m}\hat{\rho}\ket{n}$ with $m<n$, with the opposite blocks fixed by Hermitian conjugation.

Adding a diagonal internal Hamiltonian allows us to separate the spectra of the different off-diagonal spin blocks while leaving their internal level statistics unchanged.
For example, we can choose on-site energies $E_m=2^{7+m+S}$, which shift the spectrum in each coherence block $\hat{\rho}_{mn}$ by $-i(E_m-E_n)$.
These energy differences are distinct and separate the spectra.
Using the same spatial disorder and random perturbation coefficients in every block then gives shifted copies of the original coherence spectrum and its complex conjugate.
The zero-charge sector contains $(2S+1)L^2$ of the $(2S+1)^2L^2$ eigenvalues, so its relative contribution to the full spectral statistics is only $1/(2S+1)$ and vanishes as $S\to\infty$.
The dynamical bound becomes $\frac12\|\hat{\rho}_\varepsilon(t)-\hat{\rho}_0(t)\|_1\leq\varepsilon S/(2\Gamma)$, so with $\varepsilon=\varepsilon_0/L$, taking $S/L\to0$ also makes the dynamical difference vanish uniformly in time.

\subsection{Level repulsion as an unreliable diagnostic: implications for the Grobe–Haake–Sommers conjecture}
The driven harmonic oscillator illustrates how strong non-normality can undermine the interpretation of spectral level repulsion as a signature of chaos.
Although the analytical model has a regular spectrum and regular dynamics, the numerically obtained spectrum exhibits Ginibre-like correlations.
As shown above, this discrepancy originates from the large eigenvalue condition numbers of the Lindbladian, which strongly amplify numerical round-off errors.
%
%
In this example, spectral features commonly associated with chaos therefore arise through the amplification of numerical errors by non-normality.

This sensitivity cannot be dismissed as a finite-truncation effect that disappears upon increasing the Hilbert-space dimension.
As shown in Fig.~\ref{fig:levelstat}(c), the bulk eigenvalue condition numbers grow exponentially with $N_{\mathrm{tr}}$, making the spectrum sensitive to perturbations on an exponentially decreasing scale.
Thus, increasing the truncation to obtain meaningful bulk statistics intensifies the spectral instability.
The oscillator example consequently demonstrates a limitation of using the Grobe--Haake--Sommers conjecture as a general chaos diagnostic: Ginibre-like correlations can emerge from the spectral sensitivity of a regular model, without identifying any underlying chaotic dynamics (recall result~(\ref{eq:fidelity}) demonstrating dynamical stability).

The validity and interpretation of the GHS conjecture have also been debated in other dissipative models~\cite{villasenor2024breakdown,villasenor2025correspondence}.
In particular, studies of the open Dicke model found Ginibre-like level statistics in parameter regimes without strange attractors in the corresponding classical dynamics~\cite{villasenor2024breakdown}.
Subsequent work~\cite{mondal2025transient} emphasized the distinction between transient and steady-state chaos in dissipative systems.
Because dissipation contracts the accessible phase-space volume, chaotic behavior at short or intermediate times need not persist in the long-time limit.
That work accordingly interpreted Ginibre statistics as a signature of transient chaos, rather than necessarily of a chaotic asymptotic attractor.

The oscillator example exposes a separate obstruction to such interpretations.
Its underlying dynamics is regular, while strong non-normality allows numerical round-off to generate the spectral correlations usually associated with chaos.
Changing the time window in which chaos is sought therefore does not resolve this mismatch: the spectral signature itself can arise from perturbative sensitivity without a corresponding dynamical signature.
The controlled lattice construction further shows that the separation between spectral statistics and dynamics can also occur for explicit physical perturbations.
Together, these examples establish that level repulsion alone cannot serve as a general diagnostic of either transient or persistent dynamical chaos. In the Supplementary material~\cite{SuppMat} we complement these results with yet a third example: a 1D tight-binding lattice with incoherent imbalanced left/right hopping. Similar to the lattice model of the previous section, but for a simpler model, we numerically demonstrate a crossover from absence of level-repulsion to Ginibre statistics while the fidelity remains arbitrarily close to unity. 

Some care is needed when interpreting the separation between spectral statistics and dynamics.
As discussed around Fig.~\ref{fig:hamiltonian_crossover}, even a weak perturbation of an integrable Hamiltonian can produce Wigner--Dyson statistics in a sufficiently large system, while the perturbed and unperturbed evolutions remain close for $t\ll1/\varepsilon$.
The dynamical consequences of integrability breaking may therefore become apparent only at much later times.
Since unfolding removes the absolute spectral scale, the level statistics alone do not reveal this delay.

One might therefore wonder whether the strongly non-normal oscillator merely amplifies this Hamiltonian scenario, with its Ginibre-like spectrum anticipating chaotic dynamics that emerges only at extremely late times.
In the oscillator, however, the numerical errors responsible for the Ginibre-like statistics do not discernibly affect the steady state, so this cannot be the solution. 
The lattice construction provides a sharper answer to whether an appreciable dynamical difference has simply been postponed.
There, the perturbation $\varepsilon=\varepsilon_0/L$ changes the spectral statistics from 2D Poisson to Ginibre, while the fidelity bound~\eqref{eq:main-bound} guarantees that its effect vanishes with increasing $L$, uniformly over all initial states and all times.
Thus, the expectation values of every observable with uniformly bounded operator norm can remain arbitrarily close indefinitely, despite the different spectral statistics.
This holds even though the mixing time diverges in the thermodynamic limit.

Already for the modest system size considered in Fig.~\ref{fig:HNetcmodel}(c), the largest observed infidelity among the initial states and times tested is only of order $10^{-10}$.\footnote{This number is related to our choice $\Gamma=1$; we could equally well have chosen a larger $\Gamma$ with identical spectral statistics, but where the infidelity would have been even smaller.}
The distinction between Poisson and Ginibre statistics therefore need not correspond to any appreciable distinction in physical predictions, either immediately or after an arbitrarily long waiting time.

Invoking steady-state quantum chaos~\cite{richter2025integrability,rufo2025quantum} does not resolve this mismatch.
In the lattice construction, the diagonal spin blocks evolve exactly identically with and without the perturbation, while the affected off-diagonal coherences decay.
The slow late-time dynamics and the limiting physical states are therefore unchanged by construction, although the bulk spectral statistics differ.
Together with the insensitivity of the oscillator steady state to the numerical errors, this shows that Liouvillian bulk level repulsion cannot possibly diagnose steady-state chaos either.

Our examples exploit exponential spectral sensitivity.
Recall that Ginibre statistics are the generic expectation within the relevant symmetry sector, so a generic perturbation capable of changing the bulk statistics is expected to drive them towards Ginibre correlations.
Strong non-normality makes even extremely weak perturbations sufficient to produce this change already at very small system sizes.
Our examples thus realize a particularly pronounced version of the separation discussed around Fig.~\ref{fig:hamiltonian_crossover}(c,f); the onset of level repulsion need not identify a physically relevant time scale.

This separation can also occur for normal generators through the increasing spectral density discussed earlier.
As the local spacings shrink, also for normal Lindbladians progressively weaker perturbations can change the bulk spectral correlations, although a fixed weak perturbation generally requires larger system sizes to do so without non-normal amplification.
When the spectrum has comparable extents in the real and imaginary directions, typical bulk decay rates are much larger than the local eigenvalue spacings.
This leaves room for weak perturbations to reorganize the bulk statistics while the affected modes decay before any appreciable dynamical difference develops.
On spectral grounds alone, the only way to close this time-scale window is for the bulk to lie sufficiently close to the imaginary axis: its decay rates must become small enough that the modes survive until a perturbation capable of changing their statistics can appreciably affect the dynamics.
Thus, the narrowness of the spectrum must be assessed relative to the perturbation scale required to rearrange its local correlations.

Whether a particular physically admissible perturbation realizes this separation depends on how it couples to the bulk and slow modes.
A perturbation that predominantly affects the bulk while leaving the slow sector essentially unchanged can reorganize the spectral statistics without appreciably changing the observable dynamics.
The relevant scale is the decay time of its dynamical contribution, including any transient amplification, rather than the full mixing time.
Our lattice construction demonstrates explicitly that this separation can occur even when the mixing time diverges in the thermodynamic limit.
Consequently, the absence of exponential condition-number growth does not by itself justify using the level repulsion as a diagnostic.
Such an interpretation would require a model-specific argument excluding physically admissible perturbations that change the bulk correlations while leaving the slow dynamics essentially unchanged.

Strong non-normality makes the required confinement still more restrictive.
If the spectral reorganization requires only an exponentially small perturbation, the corresponding dynamical time scale can be exponentially long, so avoiding the separation can require the bulk to lie exponentially close to the imaginary axis.
Thus merely having a spectrum that appears narrow on microscopic scales is therefore insufficient.
So, without additional justification, a Hamiltonian-like interpretation of the spectral statistics based on spectral information alone can therefore remain true at best for approximately normal Lindbladians whose bulk spectrum lies sufficiently close to the imaginary axis.

Our results show that level repulsion cannot generally be taken as a signature of quantum chaos.
Two natural questions remain: whether uncorrelated spectral statistics imply the absence of chaos, and whether a spectral diagnostic becomes reliable for approximately normal Lindbladians with spectra confined close to the imaginary axis.
A purely Hamiltonian example sheds light on both questions.
Consider a chaotic Hamiltonian $\hat H$ with Wigner--Dyson statistics and the corresponding Lindbladian $\mathcal L=i[\hat H,\cdot]$.
This generator is normal and has a purely imaginary spectrum, with eigenvalues $\mu_{ij}=i(E_i-E_j)$.
Nevertheless, its difference spectrum does not inherit the Hamiltonian's level-repulsion: two differences can nearly coincide when two pairs of widely separated energies span almost equal intervals, without any original eigenvalues approaching one another.
In fact, it is these large separations in the original spectrum that dominate the short range correlations in the difference spectrum. 
We can without loss of generality rescale the Hamiltonian to have finite bandwidth and then the density of energy eigenvalues will be of the order of $D^{-1}$ where $D$ is the dimension of the Hilbert-space the Hamiltonian acts in.
Now, consider a fixed unfolded window around a nonzero bulk difference, containing an $O(1)$ expected number of differences.
Then among the differences in this window, the expected number of configurations that share an energy index is $O(D^{-1})$ and the number of eigenvalues involving distinct constituent energies separated by less than $D^{-1/2}$ is $O(D^{-1/2})$ and therefore both become negligible for large $D$.
Thus, with probability tending to one as $D\to\infty$, the relevant energies are separated by at least $D^{-1/2}$, much larger than the $O(D^{-1})$ Hamiltonian mean spacing in the bulk.
For such bulk energy arguments, the GUE correlation functions factorize~\cite{mehta2004random,soshnikov2000determinantal}:
\begin{align}
    \frac{
        \mathbb E\!\left[\prod_{a=1}^{q}n_H(x_a)\right]
    }{
        \prod_{a=1}^{q}\mathbb E[n_H(x_a)]
    }
    &\longrightarrow1 
    &
    &\text{if} 
    &
    D\min_{a\neq b}|x_a-x_b|&\longrightarrow\infty.
\end{align}
So, the part of the spectrum that is relevant for the statistics of differences is the same for a GUE Hamiltonian and a Poisson Hamiltonian.
The difference spectrum for a GUE Hamiltonian can thus show no level-repulsion. 
(The same mechanism yields Poisson local statistics in the higher-particle sectors of noninteracting Hamiltonians with Wigner--Dyson single-particle spectra~\cite{munoz2006spectral}.)
We confirm this numerically: in Fig.~\ref{fig:HNetcmodel}d the distribution of unfolded bulk level spacings for $\mathcal L=i[H_{\mathrm{GUE}},\cdot]$, with $D=128$ Hamiltonians $H_{\mathrm{GUE}}$ drawn randomly from the GUE is shown.
The resulting spacing distribution is indistinguishable from that of a one-dimensional Poisson process.
Thus, chaotic dynamics is compatible with the absence of level repulsion.
Here the Poisson statistics are one-dimensional because the eigenvalues lie on the imaginary axis, but we also have preliminary results showing that chaotic Lindbladians can exhibit genuinely two-dimensional Poisson statistics~\cite{Klein-KvorningUnpublished}.
An explicit demonstration is beyond the scope of the present paper and will be presented elsewhere.

\section{Discussion}

In this work, we have critically re-examined the idea that spectral level repulsion signals chaotic behavior in open quantum systems.
This includes the Grobe–Haake–Sommers conjecture, which asserts that if an open quantum system has a classically chaotic counterpart—e.g., featuring strange attractors or positive Lyapunov exponents—then the spectrum of its Lindbladian should exhibit level repulsion consistent with the Ginibre ensemble.
We have explicitly demonstrated a limitation of employing level statistics as a chaos diagnostic: even models with fully regular dynamics can display spectra exhibiting Ginibre-like level repulsion.

The driven harmonic oscillator provides a particularly simple illustration of this problem.
Its analytical spectrum and dynamics are regular, yet numerical round-off produces Ginibre-like spectral correlations.
The origin of this discrepancy is the strong non-normality of the truncated Lindbladian: bulk eigenvalue condition numbers grow exponentially with the Fock-space cutoff, allowing progressively smaller perturbations to rearrange the spectrum.
Increasing the truncation therefore intensifies the spectral sensitivity instead of removing it.
This behavior is associated with boundary-sensitive eigenmodes and a skin effect in Liouville space, without requiring the evolving physical state to approach the truncation boundary.
Meanwhile, the steady state remains insensitive to the numerical errors responsible for the apparent level repulsion.

The lattice construction makes the separation between spectral statistics and dynamics quantitative.
An explicit, physically admissible perturbation with strength $\varepsilon=\varepsilon_0/L$ changes the bulk statistics from Poisson to Ginibre.
Nevertheless, the fidelity bound~\eqref{eq:main-bound} guarantees that its effect on all observables of uniformly bounded norm vanishes with increasing system size, uniformly over all initial states and all times.
The perturbation leaves the diagonal spin blocks unchanged and acts only on the off-diagonal spin blocks that decay.
Consequently, the different spectral statistics cannot be explained by an appreciable dynamical difference that merely appears at later times.
This remains true even though the mixing time diverges in the thermodynamic limit.

Our explicit examples focus on strongly non-normal generators, where exponential spectral sensitivity makes the separation between spectral statistics and dynamics particularly pronounced already at modest system sizes.
As discussed above, however, increasing spectral density provides a similar mechanism even for normal generators.
Unless the spectral bulk is confined sufficiently close to the imaginary axis, one should therefore expect that a perturbation whose strength vanishes with system size can account for the observed level repulsion while remaining dynamically negligible.
In the absence of further model-specific justification, the spectral statistics alone provide no grounds for excluding this possibility.

Precisely how narrow the spectrum must be to exclude such a perturbation depends on both its spectral sensitivity and the strength quantified by the onset of level repulsion, as illustrated by the crossover dimension $D_c^{\rm L}$ in Fig.~\ref{fig:hamiltonian_crossover}.
These properties set the perturbation scale needed to reorganize the local correlations, which must be compared with the decay rates of the affected modes.
Greater spectral sensitivity allows weaker perturbations to produce repulsion and therefore requires tighter confinement near the imaginary axis for their dynamical effect to develop before the modes decay.
We leave a quantitative characterization of this competition, and of the spectral narrowness required to exclude this mechanism, to future work.

Our main examples establish that level repulsion can coexist with non-chaotic dynamics, but the converse inference can also fail.
For the purely Hamiltonian generator $\mathcal L=i[\hat H,\cdot]$, the Liouvillian eigenvalues are energy differences $i(E_i-E_j)$.
As discussed above, these differences have locally Poisson statistics even when the Hamiltonian has Wigner--Dyson correlations and chaotic dynamics.
Thus, uncorrelated Liouvillian eigenvalues need not imply the absence of chaos, even for a normal generator whose spectrum lies exactly on the imaginary axis.

While our main result concerns the limitations of spectral statistics as a diagnostic of quantum chaos in open systems, the role of non-normality may have broader implications for understanding their dynamics. In particular, it may be relevant to anomalous relaxation in certain Lindblad systems~\cite{zinidaric2023phantom,zinidaric2022solvable,bhat2025boom,mori2020resolving}, as well as to the Liouvillian skin effect and the associated critical slowing down near dissipative phase transitions, where relaxation times can diverge without a closing of the Liouvillian gap~\cite{haga2021liouvillian}. These connections highlight the importance of eigenvector structure, alongside eigenvalues, in relating the spectral properties of open quantum systems to their dynamical behavior.

\section{Methods}

We use slightly different procedures for bulk selection and spectral unfolding in the two models.
These differences reflect the implementation preferences of the authors performing the calculations: C.B.N. for the harmonic oscillator and T.K.K. for the lattice model and the purely Hamiltonian example.
We do not expect our conclusions to depend on the precise procedures used. 

\subsection{Oscillator bulk selection and unfolding}

For the driven harmonic oscillator~(\ref{dho}), we select the dense interior of the numerically obtained spectrum before calculating the spectral statistics.
All distances are Euclidean distances in the complex plane.
Writing $d_n^{(k)}$ for the distance from $\mu_n$ to its $k$-th nearest other eigenvalue, we rank the eigenvalues using the density score
\begin{equation}
    D_k(\mu_n)=\frac{1}{d_n^{(k)}}.
\end{equation}
The neighbour rank and retained fraction were chosen by visual inspection to isolate the bulk of the spectral cloud illustrated in Fig.~\ref{fig:levelstat}(d).
For the statistics in Fig.~\ref{fig:levelstat}\textbf{a},\textbf{b}, we use $k=10^4$ and retain approximately $40\%$ of the eigenvalues with the largest density scores.

The nearest-neighbour spacings $d_n^{(1)}$ are unfolded using a Gaussian estimate of the local spectral density~\cite{pawar2025comparative},
\begin{equation}
    \rho_{\mathrm{av}}(\mu_n)
    =
    \frac{1}{2\pi\sigma^2}
    \sum_{m=1}^{n_{\mathrm{eig}}}
    \exp\!\left[
        -\frac{|\mu_n-\mu_m|^2}{2\sigma^2}
    \right],
\end{equation}
where $n_{\mathrm{eig}}$ is the number of eigenvalues and $\sigma=4.5\bar d$, with $\bar d$ the mean nearest-neighbour spacing in the selected bulk.
We define
\begin{equation}
    u_n=d_n^{(1)}\sqrt{\rho_{\mathrm{av}}(\mu_n)},
    \qquad
    s_n=
    \frac{u_n}{|B|^{-1}\sum_{m\in B}u_m},
\end{equation}
where $B$ denotes the selected bulk.
The resulting spacings have unit mean.
Complex spacing ratios~(\ref{csr}) are calculated directly and require no unfolding.

The oscillator Lindbladian is constructed using QuTiP 5.2.1 and diagonalized through NumPy 1.26.4 and its underlying LAPACK routines.
Here, the sensitivity to finite numerical precision is itself part of the phenomenon under investigation: the numerical bulk spectrum differs from the analytical spectrum~(\ref{eq:hospec}) because round-off errors are amplified by the large eigenvalue condition numbers.

\subsection{Lattice and purely Hamiltonian bulk selection and unfolding}

For the Hatano--Nelson construction and the purely Hamiltonian Lindbladian, we use a fixed bulk-selection rule for each disorder realization.
Let $\{\mu_n\}_{n=1}^{n_{\mathrm{eig}}}$ denote the spectrum under consideration, and let $n_+$ be the number of eigenvalues with $\operatorname{Im}\mu_n>0$.
For each eigenvalue in this upper half-plane, we calculate its distance to the $\lfloor n_+/2\rfloor$-th nearest other upper-half-plane eigenvalue.
The bulk set $B$ consists of the $\lfloor n_+/2\rfloor$ eigenvalues with the smallest such distances.
Thus, we retain the denser half of the upper-half-plane spectrum.

For unfolding, let $R_{n,j}$ denote the distance from $\mu_n$ to its $j$-th nearest other eigenvalue in the complete supplied spectrum.
With
\begin{equation}
    k_{\mathrm{u}}=\left\lceil\sqrt{n_{\mathrm{eig}}}\right\rceil,
\end{equation}
the unfolded spacings are
\begin{equation}
    s_n=
    \frac{R_{n,1}/R_{n,k_{\mathrm{u}}}}
    {|B|^{-1}\sum_{m\in B}R_{m,1}/R_{m,k_{\mathrm{u}}}}.
\end{equation}
This compensates for variations in the local two-dimensional density and normalizes the bulk mean spacing to unity.
Each disorder realization is normalized separately before pooling the samples.
The complex spacing ratios use the nearest and second-nearest other eigenvalues in the complete supplied spectrum, with only their reference eigenvalues restricted to $B$.

\subsection{Numerical stabilization and precision tests}

The spectra in the off-diagonal spin blocks were calculated using double-precision.
To improve numerical stability, we remove the common scalar contribution $-2c$ before diagonalization, where
\begin{equation}
    c=g+\frac{\Gamma}{2\tanh(\lambda^{-1})},
\end{equation}
and restore this shift to the eigenvalues afterwards.
For $L=64$ and $128$, we additionally apply a diagonal similarity transformation,
\begin{align}
    (\mathcal S_\beta)_{(i,i'),(i,i')}
    &=
    \exp\!\left[
        \beta\frac{i+i'-2}{L-1}
    \right],
    \\
    \mathcal L_\pm
    &\longmapsto
    \mathcal S_\beta
    (\mathcal L_\pm+2c)
    \mathcal S_\beta^{-1}.
\end{align}
For the data in the figure, calculations use $\beta=-12$ for $L=64$ and $\beta=-40$ for $L=128$.
These transformations preserve the spectrum in exact arithmetic and change only its numerical representation.

Numerical sensitivity was assessed by varying the similarity transformation, retaining or subtracting the common scalar shift during diagonalization, and comparing independent numerical installations.
Eigenvalue comparisons were assessed using mutual, one-to-one nearest-neighbour matching, with discrepancies measured relative to the local nearest-neighbour spacing.
For the checked stabilized representations at $L=16,32,64$, the discrepancies were below $10^{-3}$ of that spacing.
For the checked realizations with $L\leq32$, we additionally repeated the diagonalizations at increased precision, up to 256 bits, finding agreement with the double-precision eigenvalues to within $10^{-3}$ of the local nearest-neighbour spacing.

For $L=128$, we tested three independent disorder realizations using eleven numerical representations, including similarity parameters $\beta=-24,-32,-40,-48,-56$, different treatments of the scalar shift, and two numerical installations.
Across all 55 pairwise comparisons, the largest changes in the disorder-averaged cumulative distributions of $|z|$ and $\arg z$ were below $3.8\times10^{-3}$ and $2.7\times10^{-3}$, respectively.
(However, a few individual perturbed eigenvalues at $L=128$ were less stable.)
The distribution comparisons therefore establish the stability of the reported spacing-ratio statistics under the tested numerical representations, without asserting the same accuracy for every individual eigenvalue.

\section*{Data availability}
The data concerning the Harmonic oscillator that support the findings of this study are available from the corresponding author upon reasonable request.

The full spectral data for the lattice model are available on Zenodo at \url{https://doi.org/10.5281/zenodo.24305157}.

\section*{Code availability}
The numerical simulations regarding the Harmonic oscillator model were performed using Python 3.12.3 and the QuTiP package~\cite{lambert2024qutip5quantumtoolbox}, available at \url{https://www.qutip.org}.
The lattice-model calculations used Python 3.8.10 with NumPy 1.24.4 and SciPy 1.10.1.
Precision tests additionally used MPLAPACK for quadruple-precision diagonalization and \texttt{mpmath} for independent arbitrary-precision checks.
The custom code used to generate the results is available from the corresponding author upon reasonable request. 

\section*{Acknowledgements}
T. K. K. acknowledges funding from the Wenner-Gren Foundations.

\section*{Author contributions}
All authors contributed to the conception and development of the project, the numerical simulations, the interpretation of the results, and the preparation of the manuscript. 
The author order does not reflect the extent of TKK’s contributions, particularly to the conceptual development of the project and the interpretation of the results.

\section*{Competing interests}
The authors declare no competing interests.


\bibliographystyle{naturemag}

\clearpage

\includepdf[pages=-]{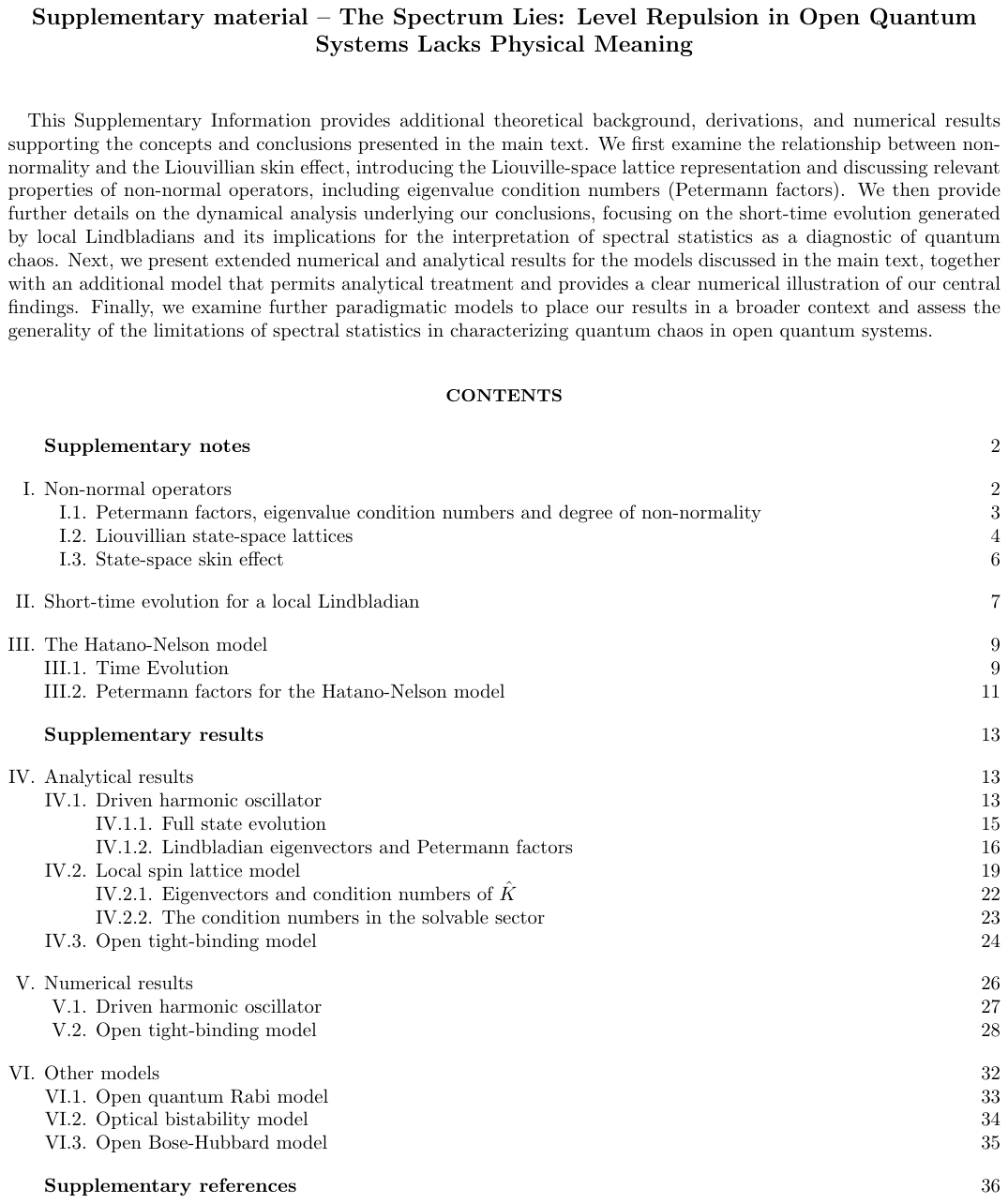}

\end{document}